# TOWARDS PERFORMANCE MEASUREMENT AND METRICS BASED ANALYSIS OF PLA APPLICATIONS

Zeeshan Ahmed

University of Wuerzburg, Wuerzburg Germany
zeeshan.ahmed@uni-wuerzburg.de

## ABSTRACT

*This article is about a measurement analysis based approach to help software practitioners in managing the additional level complexities and variabilities in software product line applications. The architecture of the proposed approach i.e. ZAC is designed and implemented to perform preprocessesed source code analysis, calculate traditional and product line metrics and visualize results in two and three dimensional diagrams. Experiments using real time data sets are performed which concluded with the results that the ZAC can be very helpful for the software practitioners in understanding the overall structure and complexity of product line applications. Moreover the obtained results prove strong positive correlation between calculated traditional and product line measures.*

## 1. INTRODUCTION

Driven by the problems raised in software industry, especially the maintenance of success rate of software products based on product line architecture. Due to the high expectations of customers for high quality product in short time and in possible minimum budget, software practitioners are now more focused on the use of product line architectures instead of single line product engineering and development. Product line is defined as the family of products designed to take the advantage of their common aspects and predicted variability [3]. For the time being product line approach was considered as the good solution to adopt but with the passage of time some feature interaction, process management, product integration and composition problems are identified [16] which are currently present without any solid solution. These problems requiring quantitative management based comprehensive solutions are the major responsible reasons of getting unsatisfactory results. A number of measurement analysis based solutions have been developed but still the problems are alive, discussed in section 2 of this research paper.

In order to provide quantitative management based solutions to the effected products, it is mandatory to first analyze all the components and features to identify the variation points. Because without knowing the actual disease it's very hard to cure and the wrong diagnose may lead to a disaster. Moreover, quantitative management is actually the way towards the successful project management. It applies measurement analysis to analyze product, predict the minimum possible cost and time required for software application development [18]. Measurement analysis can be applied to all artifacts of the software development life cycle to measure the performance and improve the quality. Measurement analysis allows stating quantitative organizational improvement objectives and decomposes them down to measurable software process and product characteristics. Moreover measurement analysis allows characterizing and controlling in a quantitative manner to achieve the organization's business goals by mapping them to software and measurement goals by using such approaches as for example GQM [20]. This can help project managers to track their improvements and mitigate potential risks.

In this research paper I discuss some already proposed solutions in section 2 to identified aforementioned problems and support software practitioners with effective and efficient quantitative management based approach in section 3 and 4. Moreover I perform empirical evaluation to evaluate the feature, functionality and strength of proposed solution in section 5 using real time data sets.

## 2. RELATED RESEARCH

### 2.1. Review Design

I decided to review only the most relevant and up-to-date literature with respect to the goal and scope of this research. I started looking for explicitly available relevant literature from different sources like internet, proceedings of some software engineering conferences and journal publications. As the outcome, I resulted with almost more than 100 relevant research papers, which created a review population. Therefore, I defined review criteria to filter the research papers based on the conclusive outputs obtained from abstract, summary and conclusion. While reading the abstract I was looking for the keywords like software product line, quantitative management, preprocessed code, analysis, visualization, measurement and variability. After the implementation of defined review criteria, 8 research papers were selected for detailed review. During detailed review, I put emphasis on the validity of the proposed solutions in literature. I have briefly described the methods discussed by the authors in the selected literature and up to what extent they can be used to solve targeted problem.

### 2.2. Reviewed Literature

I have reviewed Relation-based approach [6], Static Scanning Approach [10] and Columbus [14] as the methodologies to quantitatively analyze the C++ source code. Then I have reviewed TUAnalyser & Visualization [19] and CodeCrawler [13] to produce visualization of obtained results from quantitative analysis. Then I have reviewed Polymorphism Measure [1], Assessing reusability of C++ Code: [5] and Evaluation of Object Oriented Metrics: [7] methodologies to perform measurement analysis.

- Relation-based approach is used for simplifying the process of metrics extraction from the C++ object oriented code [6]. In this approach relations are designed like prolog clauses to describe the language entity relationship and to increases the code readability, optimization and search speed.

- Static scanning approach breaks non-preprocessed C/C++ code file into a series of lexical tokens and then matches already existing vulnerable patterns with series of tokens [10]. Furthermore the approach was validated with the implementation of ITS4 .

- Columbus is a reverse engineering framework to extract the facts from preprocessed C++ code by acquiring the project information is indispensable way [14]. The major strengths of this approach are, this method can be used for reverse-engineering purposes to study the structure, behavior and quality of the C++ code and can also be used to track the evolution of design decisions between the architectural level and the implementation level of a software system written in C++ [15].

- TUAnalyser & Visualization approach is used to visualize the extracted information from C++ source code by first storing the extracted information into RSF format which then passed as input to Graphviz to produce visual output [19].

- Code Crawler is used to visualize the C++ object oriented source code semantics in lightweight 2D & 3D and polymeric views by using a tool called FAMIX [13].

- Polymorphism measure is used to identify the software reliability problems in the early stages of software development life cycle [1]. This measurement approach has been categorized into two categories Static polymorphism and Dynamic polymorphism. Static polymorphism based on the compile time and dynamic is based on the runtime binding decisions. Furthermore five metrics are initiated to combine the early identified polymorphism forms with inheritance relationship.

- Assessing reusability of C++ Code is a method for judging the reusability of C++ code components and assessing indirect quality attributes from the direct attributes [5]. This method has been divided into two phases. First phase is used to identify and analytically validate a set of measurements for assessing direct quality attributes. The second phase identifies and validates a set of measurements for assessing indirect quality attributes. Moreover, a set of relations is also provided which maps directly measurable software quality attributes to another set of quality attributes that can only be measured indirectly.

- Evaluation of Object Oriented Metrics approach is used to analyze the measures and their relationship with each other from object oriented code metrics defined by Chidamber and Kemerer [17] and fault-proneness across three different versions of this target application [7]. Proposed approach concluded with the result, that none of the object oriented metrics performs better than LOC in anticipating fault-proneness.

### 2.3. Review Remarks

The reviewed literature doesn't provide much information about the comprehensive management for product line management, analysis of generic software and visualization for obtained results of measurement analysis from generic software components made in C++. Discussed related research work does, however, help and provide some approaches in analyzing the C++ code, visualizing the data and calculating the measures. Yet, none of the reviewed research directly deals with the target solution of this research.

## 3. PROPOSED SOLUTION

I propose and present a quantitative management based approach consisting of three main components .i.e., Analysis, Visualization, and Measurement shown in Fig 1. During analysis the preprocessed source code of product line based application is analyzed to identify the software single and product line source code characteristics, then in measurement several traditional and product line measures are calculated to identify the behavior and the rate of complexity and then in the end qualitative visualization support is provided to obtained results. To provide a comprehensive solution to the software practitioners, based on above discussed three components in analyzing the software product preprocessed source code, identifying software level complexities and variabilities, measuring performance by calculating source code metrics I have proposed a solution called Zeeshan Ahmed - C++ Prepreprocessed source code analyzer (ZAC) [23] [24] [25].

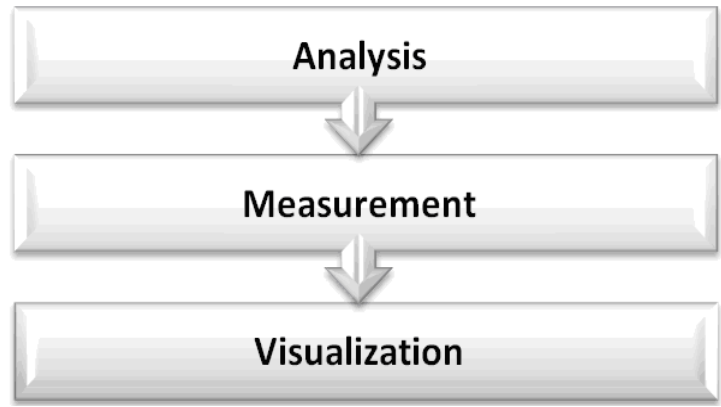

Fig 1. Measurement Analysis and Visualization

To fulfil the desired jobs and obtain required result the conceptual model of ZAC is divided in to five main components .i.e., C++ Source Code, ZAC-Analyzer, ZAC-Data Manager, ZAC-Measurer, and ZAC-Visualizer which works in cyclic fashion as shown in Fig 2.

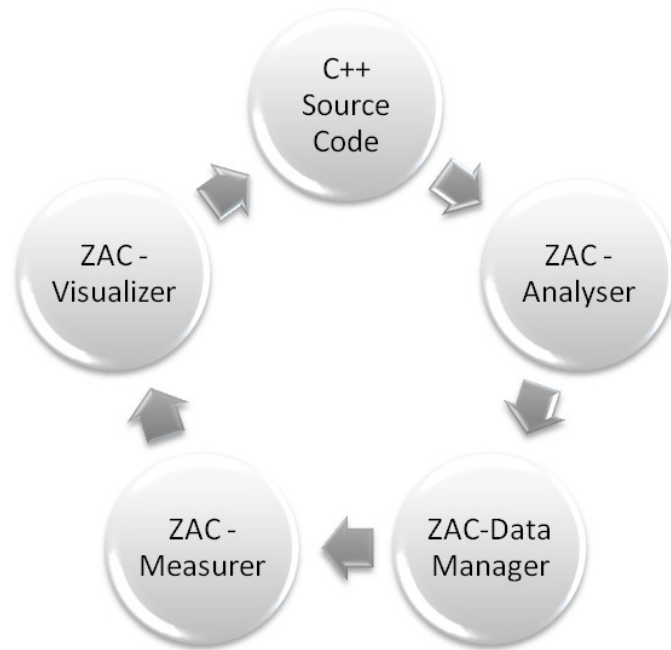

Fig 2. ZAC Conceptual Model

In the first component preprocessed source code of software product written in C++ programming language is treated as input context. In second component internal source code characteristics .i.e., namespace, includes, macros, class, methods, functions, declarations, expressions and conditions are analyzed. In third component the resultant information is stored and managed in a relational data base called ZAC-Data Manager. In fourth component ZAC-Measurer using Goal Question Metrics (GQM) [20] will calculate source code metrics. In the last component, as the last step ZAJ-Visualizer will produce visualization of obtained results in different 2D and 3D diagrams e.g. graphs, line charts, bar charts and tree map with respect to the context and semantic.

## 4. ZAC

To empirically evaluate the real time strength of proposed solution, I have implemented the proposed conceptual design of ZAC as software application. ZAC is implemented using open source and freely available development tools and technologies as shown in Fig 3.

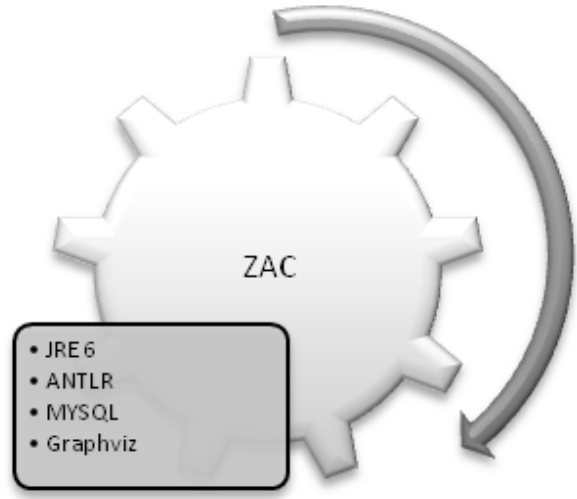

Fig. 3 ZAC - Involved Technologies

According to the scope of this research, application is capable of treating preprocessed C++ source code as an input. The designed and implemented system sequence model of application is consists of six main components .i.e., ZAC- Source Code Analyser, ZAC-Semantic Modeler, ZAC-Data Manager, ZAC-Measurer, ZAC-Visualizer and ZAC-Editor as shown in Fig 4.

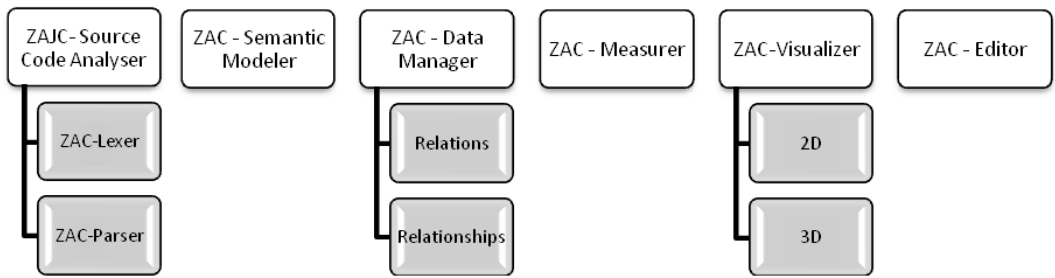

Fig 4. ZAC System Sequence Design

Preprocessed C++ source code is given to the application as an input. ZAC-Analyser first treats the input source code and analyzes the source code. To analyze preprocessed source ZAC-Analyser is divided into further two internal sub-components .i.e., ZAC-Lexer and ZAC-Parser. In ZAC-Lexer the whole application is divided into possible number of lexical tokens where as in ZAC-Parser parsing is performed using generated lexical tokens to understand and validate the syntax and semantic of the input preprocessed source code with respect to the parser rules based on grammar of used programming language used in application development. The resultant output of ZAC-Analyser consisting of the information about the total number of artifacts, classes, components, control flows, decisions, defines, directives, parameters, exceptions, expressions, features, headers, macro expressions and namespaces is used by ZAC-Semantic Modeler, which will generate a semantic based object oriented preprocessed source code model as shown is Fig 5.

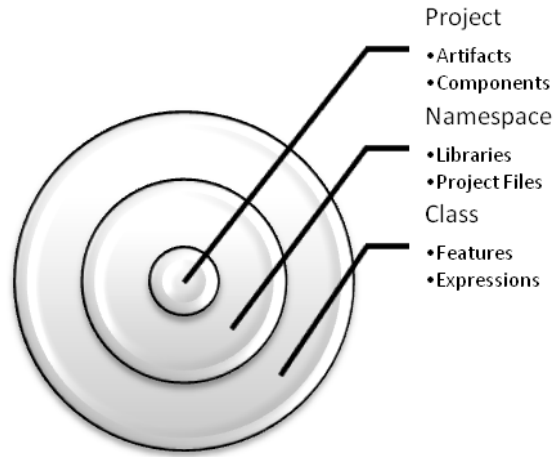

Fig 5. ZAC Object Oriented Preprocessed Source Code Model

This model is further stored in database in the form of relations and relationships using ZAC-Data Manager. Moreover generated semantic based object oriented preprocessed source code model is also used by ZAC-Measure to calculate traditional and product line measures metrics. The output of the ZAC-Analyzer and ZAC-Measures is visualized in visual diagrams .i.e., tree maps, tree graphs and pie charts in two dimensional effects using ZAC-Visualizer. This all process above discussed process is based on automatic operational processing but as the last component ZAC-Editor provides several different options to manually operate analyze the software characteristics and calculate metrics.

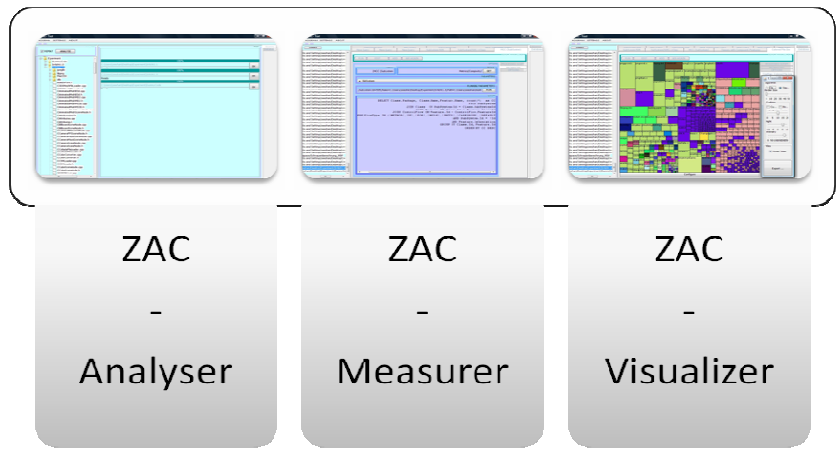

Fig. 6  ZAC Screen Shots

# 5. EMPIRICAL EVALUATION

The main goal of this empirical evaluation is to evaluate the feature, functionality and strength of ZAC by performing experiments using real time software product line application developed in C++ programming language as an input context. I have to first analyze software preprocessed source code characteristics, then calculate the traditional and product line measures and visualize the results. Moreover using the outcome of this experiment I try to observe and present the correlation between traditional and product line applications.

## 5.1. Context of empirical evaluation

Two available different source code versions (1.0 & 1.1) of an open source product line application Irrlicht are used as the context for the evaluation. Irrlicht is a cross-platform high performance real time 3D application (engine) developed in C++ language [9].

## 5.2. Defined measures for empirical evaluation

It is quite difficult and time consuming to calculate all the traditional and product line measures, so, I have defined some traditional and product line measures which will be calculated from input preprocessed source code during the empirical evaluation. I have selected three relevant measures from over all currently available traditional and product line measures. The information about the selected traditional and product line measure is given in Table I and Table II.

TABLE I
DEFINED TRADITIONAL MEASURE

| Traditional Measures | Definition | Comments |
|---|---|---|
| CLD (Class Leaf Depth) | CLD is used to measure maximum number of levels in the hierarchy of classes that are below the class. | I expect that the CLD helps in the indication of fault proneness. Because a class with many ancestors is more likely to be fault-prone than a class with few ancestors [12]. So Increase in the number CLD can cause the increase in the complexity of the software which then can be resulted in more fault proneness. |
| NOC (Number of children) | NOC is used to measure the number of direct descendent for each class. | NOC helps in the indication of the fault proneness. Lower the rate of fault proneness if higher the rate of NOC [21]. |
| DIT (Depth of inheritance) | DIT is used to measure the ancestors of the class. | DIT helps in the indication of fault proneness. The DIT was found to be very significant in [21], it means that the larger the DIT, the larger the probability of fault-proneness in the software. |

TABLE II
DEFINED PRODUCT LINE MEASURE

| Product Line Measures | Definition | Comments |
|---|---|---|
| NIT (Namespace Inheritance Tree) | NIT is used to measure the number of ancestors of the name spaces. | NIT can also be very helpful in the indication of fault proneness. Because namespace is one of the major components of any product line software application. Every component of project like class, library etc exists inside the namespace. If the number of namespace will increase, then the number of ancestor namespaces will also be increased which will result in much complexity. So if the NIT will increase it may possible then the number fault proneness will also increase. |
| NOA (Number of Artifacts) | NOA is used to measure the number of artifacts used and to measure the direct ancestors for each artifacts. | Each artifact of system represents the each file of the system, like class and include file. If the number of artifacts will increase most probably the number of dependencies between the artifacts will also be increase, which ultimately causes the increase in complexity. So higher the number of NOA higher will be the probability of fault proneness in the product line system. |
| CIR (Class Inheritance Relationship) | CIR measures the relationship of each class with other class. | CIR becomes more complex if the number of ancestor classes will increase, especially in the case of multiple inheritances when a child class will be having more than one parent. So, CIR can be very helpful in the indication of fault proneness because if the number of CIR will increase the probability of fault proneness will also be increased. |

### 5.3. Defined visualization modes for empirical evaluation

I have defined some modes of visualization to present the obtained results using ZAC during the empirical evaluation. I have selected four modes of visualization i.e. Tree Map, Namespace Graph, Bar chart and Inheritance graph, for the visual representation of obtained results.

### 5.4. Evaluation

1) Static Source Code Analysis

During the static analysis both the versions 1.0 and 1.1 of Irrlicht are analyzed using ZAC and observed experimental units of the source code are presented in Table III. I have first performed the analysis of source code characteristics and calculate the absolute and relative improvement in the latest version 1.1 as compared to the version 1.0 of Irrlicht. After analyzing the results I have come to know that the number source code with respect to each characteristic has been decreased in Irrlicht 1.1 as compared to Irrlicht 1.0.

TABLE III
GENERAL AND PRODUCT LINE SOFTWARE CHARACTERISTIC S

| Characteristic Name | Irrlicht 1.0 | Irrlicht 1.1 | Observation | Observation |
|---|---|---|---|---|
| Software characteristics | Ver 1.0 | Ver 1.1 | Absolute Improvement in Ver 1.1 | Relative Improvement in Ver 1.1 |
| Artifacts | 776 | 698 | 78 | 10.51 % |
| Namespaces | 8 | 7 | 1 | 12.50 % |
| Components | 561 | 482 | 79 | 14.08 % |
| Decisions | 703 | 445 | 258 | 36.70 % |
| Define Macros | 609 | 447 | 162 | 26.60 % |
| Pragma Directives | 11 | 10 | 1 | 9.09 % |
| Macro Expressions | 402 | 276 | 126 | 31.34 % |
| Classes | 333 | 207 | 126 | 37.84 % |
| Include | 1027 | 532 | 495 | 48.20 % |

2) Traditional and product line measures based analysis

I have calculated the already defined traditional and product line measures of Irrlicht 1.0 and 1.1, results are presented in Table IV and Table V.

TABLE IV
TRADITIONAL MEASURE

| Traditional Measure | Irrlicht | Irrlicht | Fault Proneness | Fault Proneness |
|---|---|---|---|---|
| Coupling | Ver 1.0 | Ver 1.1 | Absolute Improvement | Relative Improvement |
| CLD | 66 | 21 | 45 | 68.18 % |
| DIT | 232 | 145 | 87 | 37.5 % |
| NOC | 64 | 21 | 43 | 67.18 % |

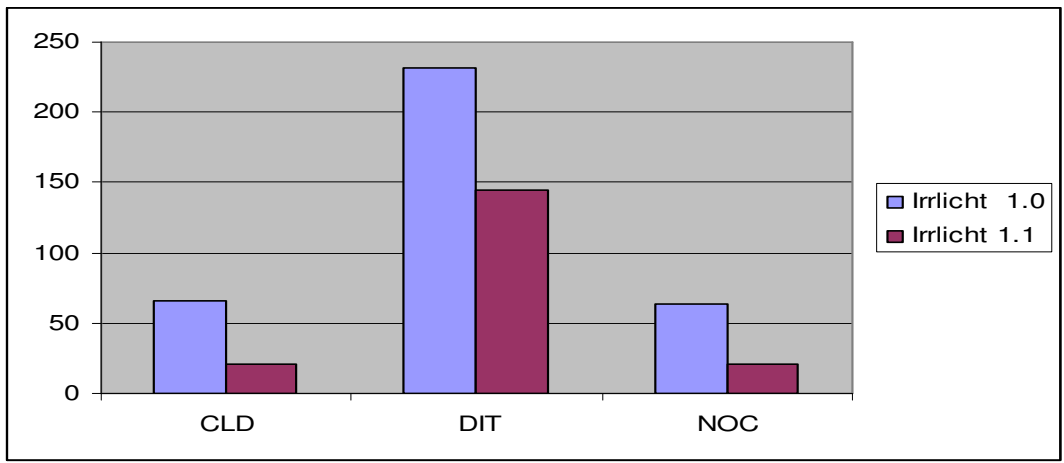

Fig. 7: Analysis Traditional Measure

TABLE V
PRODUCT LINE MEASURE

| PL Measure | Irrlicht | Irrlicht | Fault Proneness | Fault Proneness |
|---|---|---|---|---|
| Coupling | Ver 1.0 | Ver 1.1 | Absolute Improvement | Relative Improvement |
| NIT | 7 | 6 | 1 | 14.28 % |
| NOA | 783 | 704 | 79 | 10.08 % |
| CIR | 160 | 97 | 63 | 39.37 % |

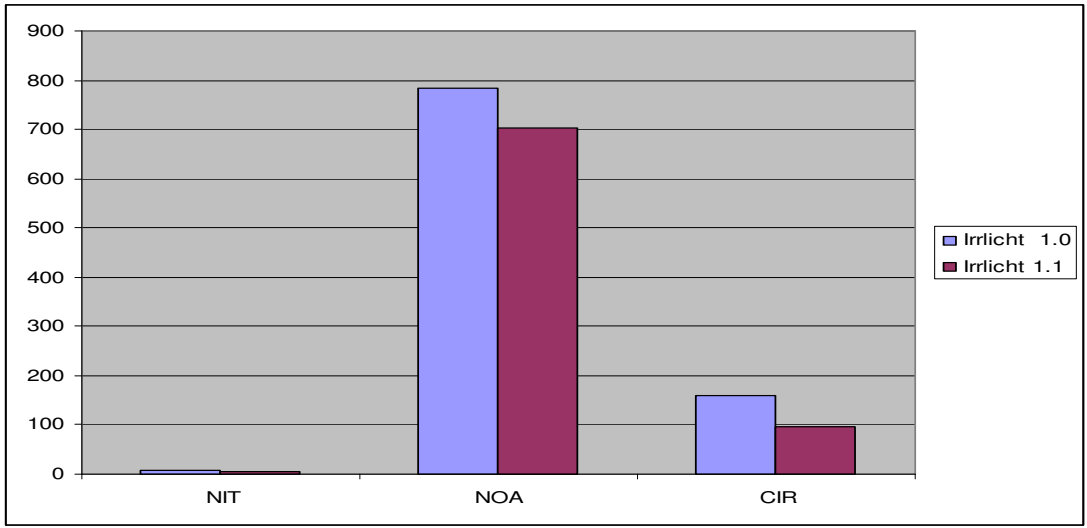

Fig. 8: Analysis Product line measure

After analyzing the results by calculating the absolute and relative improvement of fault proneness in Irrlicht 1.1 as compared to Irrlicht 1.0, I have identified that the traditional and product line measures based resultant values are decreased in Irrlicht 1.1 as compared to Irrlicht 1.0. If we see the results in Table 1V then we come to know that the number of CLD, DIT and NOC measures have been decreased in the Irrlicht 1.1 with respect to the Irrlicht 1.0, which shows that the number of fault proneness is decreasing as shown in Figure 7. Furthermore If we observe the results in Table V then we will come to know that the number of NIT, NOA and CIR measures have been decreased in the Irrlicht 1.1 with respect to the Irrlicht 1.0, which shows the decrease in the number of fault proneness shown in Figure 8.

3) Visualization of software characteristics

I have presented defined visual diagrams of the results based on the static source code analysis of the Irrlicht 1.0 and 1.1. These diagrams present some of the whole results of visualization produced by the ZAC. This visual representation can be helpful for the software practitioners in observing the overall structure and complexity of relationships between attributes namespaces and classes etc.

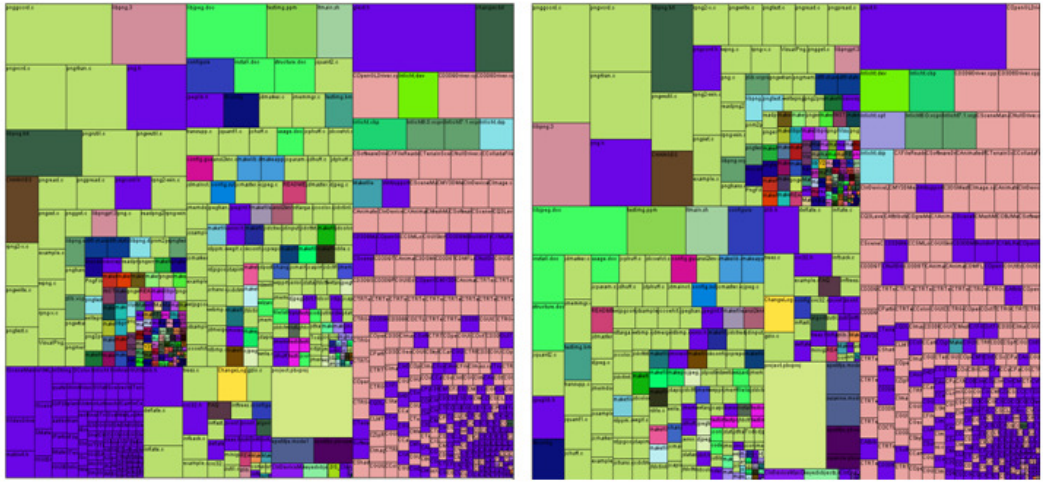

Fig 9. File Tree Map

Fig. 9 File Tree Map presents the number of files used in Irrlicht 1.0 in Figure 9 (a) and Irrlicht 1.1 in Fig 9 (b), drawn in a tree map with respect to their type and size, likewise files represented in the pink color presents the CPP files (".cpp" files), green colored files represents doc files (".doc" files). Whereas the placement of each file is with respect to the association of files with each other and directory structure. This visual representation is helpful for the software practitioners in analyzing the over file structure of both the projects and can also conclude with a vital difference as well.

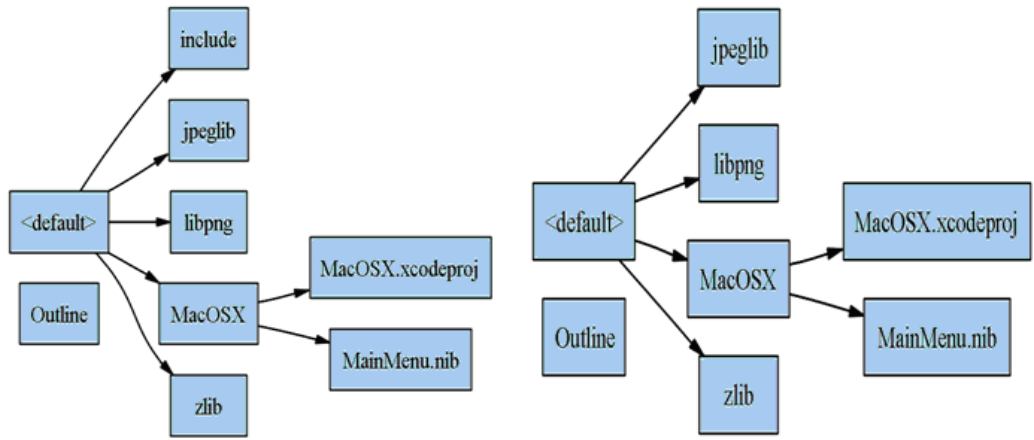

Fig 10. Namespaces Structure

Fig. 10 Namespace Structure presents the graph of namespaces used Irrlicht 1.0 and Irrlicht 1.1. As shown in Fig. 10 (a), there are 8 namespaces, default namespace behaving as the parent namespace for his children namespaces include, jpeglib, libpng, MacOSX and zlib, furthermore MacOSX is also behaving as the parent to his child namespaces MacOSX.xcodeproj and MainMenu.nib. Whereas shown in Fig. 10 (b), there are 7 namespaces in over all project Irrlicht

1.1, default namespace behaving as the parent namespace for his children namespaces jpeglib, libpng, MacOSX and zlib, further more MacOSX is also behaving as the parent to his child namespaces MacOSX.xcodeproj and MainMenu.nib This visual representation is helpful for the practitioners to analyze the overall structure of the namespaces used in the project; moreover software practitioners can also take advantage in analyzing the complexity in namespace relationship.

Fig. (a) – Bar Chart of Irrlicht 1.0          Fig. (b) – Bar Chart of Irrlicht 1.1

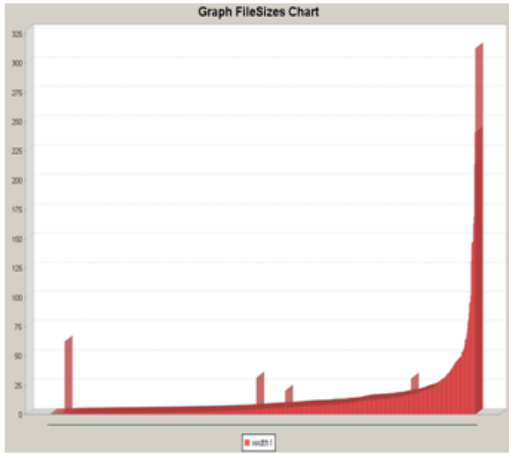     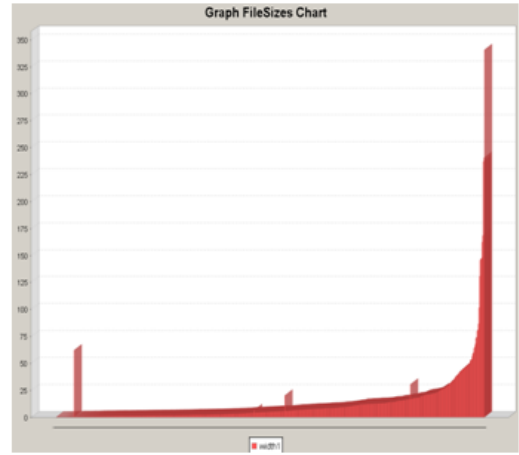

Fig  11. Bar Chart Analysis of Artifacts

Fig. 11 File bar Chart presents the total number of artifacts used in Irrlicht 1.0 and Irrlicht 1.1, drawn in a bar chart. As shown in Fig 11 (a), the bar chart of Irrlicht 1.0, consists of 775 red lines where as shown in Fig 11 (b) the bar chart consists of 698 red lines; each red line is representing an artifact and the length of each red line represents size of respective artifact used in the Irrlicht 1.1. This kind of visual representation is very helpful for the overall source code analysis. Because some time even the rate of increase or decrease in some source code elements with respect to class or namespace can also play a vital role in increasing or decreasing the complexity e.g. cohesive or coupled code.

Fig. (a) – Inheritacne Relationships Irrlicht 1.0          Fig. (b) – Inheritance Relationships of Irrlicht 1.1

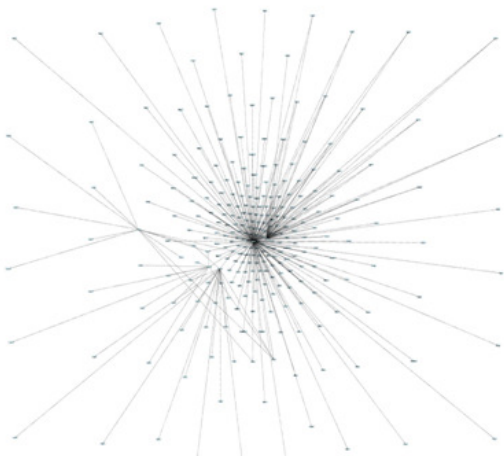     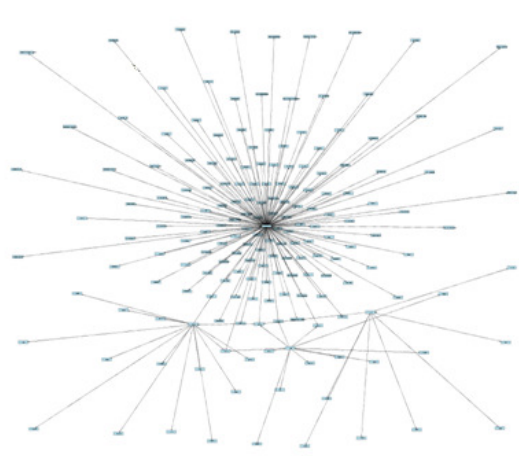

Fig 12. Inheritance Relationship

Complexity graphs based on all the used "include" in Irrlicht 1.0 (1026) and Irrlicht 1.1(532) along with their relationship with each other with respect to the namespace are presented in Figure 12 (a) and (b). This visualisation can help software practitioners in understanding the over inherited structure of applications.

## 5.5. Analysis on Results

First of all preprocessed source code of Irrlicht 1.0 & 1.1 was analyzed; the outcome of analysis was concluded with information about the decrease in each characteristic of Irrlicht version 1.1 as compared to version 1.0. Then traditional and product line measures of both the versions 1.0 and 1.1 of Irrlicht were calculated, which resulted with the information, that the number of traditional as well as the product line measures has been decreased in Irrlicht version 1.1 as compared to version 1.0.

As the last step, the correlation [2] between traditional and product line measures was calculated. This resultant correlation +0.93 show the strong correlation between calculated traditional and product line measures. As the resultant value is positive "+", so I can say higher the rate of traditional measures higher will be the rate of product line measure. Furthermore, I have also produced the visualization of some source code characteristics in graph, bar char and tree maps, which helps the software practitioners in understanding the overall structure and complexity of product line applications.

## 5.6. Limitations

During the static analysis of Irrlicht 1.0 & 1.1 some experimental units were dropped and not considered, because currently available version of ZAC is not capable of completely resolving all the experimental units of Irrlicht 1.0 and 1.1.

## 6. CONCLUSIONS

The aim of this research work was to provide the effective support to software practitioners in quantitatively managing the software product line applications by analyzing software preprocessed code characteristics, measuring complexity which may indicate the potential reliability and maintainability problems in delivered software [11] and producing visualization of results. To achieve the aforementioned research goals I have discussed some already proposed approaches by some other authors and own quantitative analysis based solution. In the end I have performed empirical evaluation to evaluate the feature, functionality and strength of ZAC by performing experiments using real time data set and concluded with results that ZAC can be very helpful for the software practitioners in understanding the overall structure and complexity of product line applications. Moreover I have also proved using obtained results in empirical evaluation that there is a strong positive correlation between calculated traditional and product line measures.

**Author Bibliography**

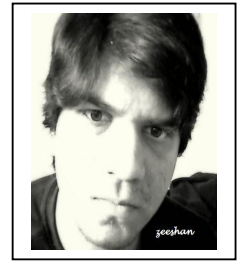

Zeeshan Ahmed is a Research Scientist, presently working in the Department of Bioinformatics- University of Wuerzburg Germany. He has on record more than 12 years of University Education and more than 7 years of professional experience of working within different multinational organizations in the field of Computer Science with emphasis on software engineering of product line architecture based artificially intelligent systems. He also has more than 4 years experience of teaching as lecturer and supervising research thesis to graduate and undergraduate students in different institutes and universities.